\documentclass[preprint,10pt,3p]{elsarticle}

\usepackage{graphicx}
\usepackage{latexsym}
\usepackage{amsmath}
\usepackage{amssymb}
\usepackage{amsfonts}
\usepackage{color}
\usepackage{listings}
\usepackage{mathrsfs}
\usepackage{comment}
\usepackage{hyperref}
\usepackage{theoremref}
\usepackage{ntheorem}
\usepackage{dsfont}
\usepackage{bm}
\usepackage{lscape}
\usepackage{float}

\newtheorem{theorem}{Theorem}

\newtheorem{defn}{Definition}

\newtheorem{remark}{Remark}

\newcommand{\z}{{\mathbb Z}}

\newcommand{\R}{\mathbb{R}}
\newcommand{\Prob}{\mathbb{P}}

\newcommand{\E}{\mathbb{E}}
\newcommand{\Normal}{\operatorname{Normal}}
\newcommand{\DExp}{\operatorname{DExp}}
\newcommand{\GH}{\operatorname{GH}}

\newcommand{\MX}{\operatorname{MX}}

\newcommand{\CGMY}{\operatorname{CGMY}}

\newcommand{\Poisson}{\operatorname{Poisson}}

\numberwithin{equation}{section}

\definecolor {grey}{rgb}{0.2,0.2,0.2}
\definecolor {dkgr}{rgb}{0,0.3,0}

\journal{NULL}

\begin{document}

\begin{frontmatter}

\title{ L\'evy Processes For Finance:\\ An Introduction in {\tt R}}

\author{D.J. Manuge}

\begin{abstract}
This brief manuscript provides an introduction to L\'evy processes and their applications in finance as the random process that drives asset models. Characteristic functions and random variable generators of popular L\'evy processes are presented in {\tt R}.
\end{abstract}

\end{frontmatter}

\section{Introduction}

Over the past few decades analysis has shown that market data is inconsistent with some of the underlying assumptions of the Black-Scholes model.  For example, if asset price changes remain large while time periods shrink, one cannot assume that prices are continuous. Under this observation, Cox and Ross assume prices follow a pure jump process so that at any timestep the asset price results in a positive jump or negative drift \cite{coxross}. Expanding on this theory, Merton introduced jumps superimposed on a continuous price process, whose parameters can be chosen to account for fat-tail returns observed in real markets \cite{1999866}. Additionally, practitioners require models with accurate implied volatility surfaces for risk management. \footnote{Known as the {\sl smile effect} of volatility, due to its graphical portrayal of happiness.} While the surface can be explained without a jump model, the {\sl smile} becomes augmented for short maturity options, which is better described by the presence of jumps \cite{tankov2003financial}. \footnote{In non-jump models, as $T \to 0$, $\ln(S_t/S_{t-1}) \sim \Normal(\mu, \sigma)$, when in reality this is not the case.} Moreover, distributions of returns exhibit skewness and leptokurtosis. Clearly, as more information about market structure is understood, improvements to parametric forms of asset prices are neccessary. For the class of finite jump models, Kou extends Merton's model by allowing jump sizes to follow an asymmetric double exponential distribution \cite{kou}. The model accounts for both the volatility smile and asymmetric leptokurtic returns. Most financial models fall into the class of infinite jump models, where many developments have been made. In 1987, Madan and Seneta suggest that increments of log-prices follow a (symmetric) variance gamma (VG) distribution, while providing statistical evidence using Australian stock market data \cite{madan1986chebyshev, 1990madan}. \footnote{Later extended to the asymmetric variance gamma model in \cite{Madan98thevariance}} The VG distribution is a special case of the generalized hyperbolic (GH) distribution, which was initially proposed by Barndorff-Nielson to model the grain-size distribution of sand as it travels from a source to a deposit \cite{barn1977}. Other cases of the GH distribution have been considered for asset price modelling; Eberlein and Keller suggest that increments of log-prices follow a hyperbolic distribution \cite{Eberlein95hyperbolicdistributions}. Alternatively, Barndorff-Nielson propose the normal inverse Gaussian (NIG) distribution for log-prices in \cite{barndorff1995normal}. Finally, Eberlein and Prause extend their focus to the entire class of GH distributions, providing a thorough statistical treatment as well as applications to stochastic volatility \cite{Eberlein98thegeneralized, prause1999generalized}. Carr, Geman, Madan, and Yor proposed the (generalized) tempered stable process. \cite{Carr00thefine}. Coined the CGMY process, it coincides with the L\'evy measure of a non-normal $\alpha$-stable L\'evy process ($\alpha \in (0,2)$) multiplied by an exponential factor. Another infinite jump model, which arose from the theory of orthogonal polynomials, is given by the Meixner process \cite{schoutensmeix, teugels}. All of the processes mentioned thus far share a common trait; they are L\'evy processes. It is obvious that by modelling the asset dynamics with a general L\'evy process, a large range of special cases are accessible for consideration. \footnote{Which include (but are not limited to) the financial cases mentioned above.} Not only does this general class of process sufficiently mimic the implied volatility surface of real data, but as mentioned they can be parametrized to exhibit skewness, kurtosis, an absence of autocorrelation in price increments, finite variance, aggregational normality, and have an ability to change discontinuously \cite{papant, papapantoleon2006, raible2000levy}. 

\section{L\'evy Processes}
\begin{defn}[Stochastic Process] 
A stochastic process $X$ on a probability space $(\Omega, \mathcal{F}, \mathbb P)$ is a collection of random variables $(X_t)_{0 \leq t < \infty}$.
\end{defn}
If $X_t \in \mathcal F_t$, the process $X$ is said to be adapted to the filtration $\mathcal{F}$, or equivalently, $\mathcal F_t-measurable$ (\cite{protter}). Stochastic processes have been studied in economic, financial, actuarial, physical, biological, and chemical applications (\cite{protter}, \cite{kyprianou2006introductory}, \cite{gichman1972gihman}, \cite{gihman-skorokhod}, \cite{karatzas-shreve}). In finance, their primary use is to represent the price component of some asset. Arguably, the most well-known stochastic process is Brownian motion, which describes the movement of particles suspended in a fluid (\cite{ein},\cite{karatzas-shreve}). 
\begin{defn} [Brownian Motion] \label{3}
Standard Brownian motion $W=(W_t)_{0 \leq t < \infty}$ satisfies the following three properties:\\
\noindent (i) $W_0=0$\\
(ii) $W$ has independent increments: $W_t-W_s$ is independent of $\mathcal{F}_s,$  $0 \leq s<t< \infty$\\
(iii) $W_t-W_s$ is a Gaussian random variable: $W_t-W_s \sim N(0,t-s)$ $\forall$ $0 \leq s < t < \infty$
\end{defn}
Property $(ii)$ implies the Markov property (i.e. conditional probability distribution of future states depend only on the present state). Property $(iii)$ indicates that knowing the distribution of $W_t$ for $t \leq \tau$ provides no predictive information about the process when $t > \tau$. Another well-known stochastic process is (\cite{karatzas-shreve}), 
\begin{defn} [Poisson Process] \label{poiss}
A Poisson process $N=(N_t)_{0 \leq t < \infty}$ satisfies the following three properties:\\
\noindent (i) $N_0=0$\\
(ii) $N$ has independent increments: $N_t-N_s$ is independent of $\mathcal{F}_s,$  $0 \leq s<t< \infty$\\
(iii) $N$ has stationary increments:  $P(N_t-N_s \leq x)=P(N_{t-s} \leq x),$    $ 0 \leq s<t< \infty $ 
\end{defn}
In reality, stochastic differential equations formulated with only Brownian motion or the Poisson process may be ineffective in describing the complex dynamics of an evolving system. Although their path structures appear different in simulation, note the similarity in their definitions. By combining their common properties, a general process can be established. \footnote{This is merely one of many results attributed to Paul L\'evy (1886-1971) who produced foundational results in the calculus of probabilities; at a time when no mathematical theory of probability was available. His doctoral advisors were Jacques Hadamard and Vito Volterra.}
\begin{defn} [L\'evy Process] \label{levy}
Let $L$ be a stochastic process. Then $L_t$ is a L\'evy process if the following conditions are satisfied:\\
\noindent (i) $ L_0=0$\\
(ii) $L$ has independent increments: $L_t-L_s$ is independent of $\mathcal{F}_s,$  $0 \leq s<t< \infty$\\
(iii) $L$ has stationary increments:  $\Prob(L_t-L_s \leq x)=\Prob(L_{t-s} \leq x),$    $ 0 \leq s<t< \infty $ \\
(iii) $L_t$ is continuous in probability: $\lim_{t \to s} L_t = L_s$
\end{defn}
\begin{remark}
Keller shows that condition (iii) follows from (i) and (ii), and thus can be omitted (\cite{keller1997realistic}).
\end{remark}
A random variable $L$ is said to have an infinitely divisible distribution if there exists a sequence of i.i.d random variables such that $L \overset{d}{=} L_{1,n}+ ...+ L_{n,n}$ for each $n \in \z^+$. It follows that if $L$ is a L\'evy process, it has an infinitely divisible distribution for each $t>0$ (\cite{kyprianou2006introductory}). \footnote{The converse holds as well; all infinitely divisible distributions can be represented as a L\'evy process.} An expression exists which fully characterizes the properties stated, giving a parametric structure to this process' seemingly qualitative definition.
\begin{theorem} [L\'evy-Khintchine formula] Suppose that $\mu \in \R$, $\sigma \geq 0$, and $\nu$ is a measure concentrated on $\R / \{0\}$ such that $\int_\R \min(1,x^2) \nu(dx) < \infty$. A probability law $\eta$ of a real-valued random variable $L$ has characteristic exponent $\Psi(u):= - \frac{1}{t} \log \E [e^{iuL_t} ]$ given by,
\begin{align}
\Phi(u;t)=\displaystyle\int_\R e^{i u x} \eta(dx) = e^{- t\Psi(u)} \hspace{5mm} \text{for} \hspace{5mm} u \in \R,
\end{align}
iff there exists a triple $(\mu,\sigma,\nu)$ such that,
\begin{align}
\Psi(u)=i\mu u+\frac{1}{2} \sigma^2 u^2 + \displaystyle\int_\R (1-e^{i u x} + i u x \mathds{1}_{(|x|<1)}) \nu(dx)
\end{align}
for every $u \in \R$.
\end{theorem} 
Furthermore, it follows from the above theorem that there exists a probability space where $L=L^{(1)}+L^{(2)}+L^{(3)}$; $L^{(1)}$ is standard Brownian motion with drift, $L^{(2)}$ is a compound Poisson process, and $L^{(3)}$ is a square integrable martingale \footnote{A martingale is a process $L$ whose conditional expected value at  $L_{n+1}$ is equal to $L_n$.} with countable number of jumps of magnitude less than 1 (almost surely). The characteristic exponent of the process is the same as in the L\'evy-Khintchine formula. This result is more formally known as the L\'evy-It\^{o} decomposition (\cite{protter}). In canonical form,
\begin{align} \label{itodecomp}
L_t = \mu t + \sigma W_t + \int\limits_0^t \int\limits_{|x| \geq 1} x \eta^L(ds,dx) + \int\limits_0^t \int\limits_{|x|<1} x(\eta^L - \nu^L)(ds,dx).
\end{align}

\section{Simulating Assets with L\'evy Processes}

The path structure of any L\'evy process is uniquely defined by its triplet $(\mu, \sigma, \nu)$. Many popular asset pricing models and their associated driving processes have known triplets (\cite{Papapantoleaon2008}, \cite{kyprianou2006introductory}). This section discusses examples of L\'evy processes that are used in finance to model assets. In the cases which follow we assume the price process takes the form, \\
$$
S=\exp(L_t)
$$
where $L_t$ is a L\'evy process. In this paper, we do not discuss details regarding the risk-neutrality of the asset. However, we note that the asset must be mean-corrected for it to be risk-neutral. One way this can be achieved is by multiplying the above expression by an exponential damping factor which can be solved by knowing the distribution of the asset (i.e. the choice L\'evy process)  (\cite{raible2000levy}). In the code that presents the characteristic functions of L\'evy processes below, risk-neutrality is applied. The "fAsianOptions" and "Bessel" {\tt R} packages should be installed and loaded prior to running the scripts in this section. \\

\noindent  \textbf{Brownian Motion:} Suppose we have mean $\mu$ and variance $\sigma^2$ with $L_1 \sim \Normal(\mu, \sigma^2)$, then the triplet is given by $(\mu, \sigma^2, 0)$. The density $\eta_{L_1}$ is given by,
\begin{align}
\eta_{L_1} (x) = \frac{1}{\sqrt{2 \pi} \sigma} e^{-(x-\mu)^2 /2\sigma^2}.
\end{align}
The canonical decomposition is given by,
\begin{align}
L_t= \mu t + \sigma W_t
\end{align}
which has a characteristic exponent of the form,
\begin{align}
\Psi(u)=\frac{1}{2} \sigma^2 u^2 - iu\mu.
\end{align}
which can be obtained via:
\begin{verbatim}
BS_CF=function(u,sigma,r,time){
drift=r-0.5*sigma^2
phi=exp(1i*drift*time*u-0.5*sigma*sigma*u*u*time)
return(phi)
}
\end{verbatim}
Random variables can be obtained to simulate the path of assets by calling:
 \begin{verbatim}
BM=function(mu,sigma, T, N) {
h=T/N
t=(0:T)/N
X=rep(0, N+1)
X[1]=0
for(i in 1:N) { X[i+1]=X[i] +mu*h+sigma*sqrt(h)*rnorm(1)}
return(X)
}
\end{verbatim}
\begin{figure}[H] 
\centerline{\includegraphics[scale=0.55]{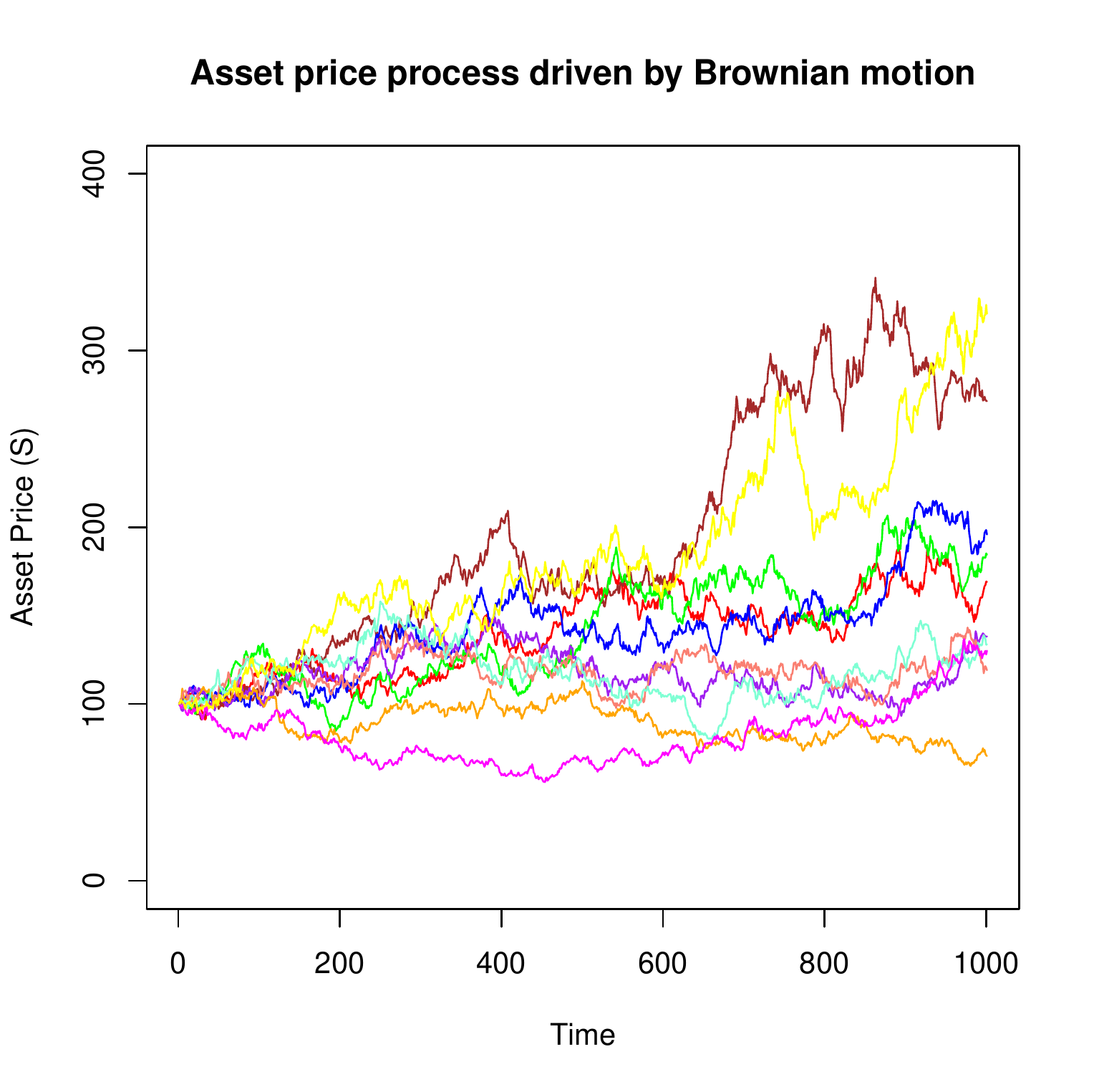}} \label{BM}
\caption{Sample paths of an asset driven by Brownian motion with $\mu=0.5$ and $\sigma=0.5$.}
\end{figure}
\noindent Log-returns are normally distributed in the BSM model; the log-price process follows a standard Brownian motion with drift (\cite{black-scholes}). Setting $\mu=0$ eliminates the drift component. \\
\textbf{Poisson Process:} Let $\lambda >0$ and $\delta_1$ be the Dirac measure for $x=1$. Suppose $L_1 \sim \Poisson(\lambda)$, then the triplet is given by $(0,0, \lambda \delta_1)$. \footnote{$\delta_1(A)=1$ when $1 \in A$, otherwise $\delta_1(A)=0$.} The density $\eta_{L_1}$ is given by,
\begin{align}
\eta_{L_1} (x) = \lambda e^{-\lambda x}.
\end{align}
The decomposition is the cumulative sum of jumps up to time $t$, which has a characteristic exponent of the form,
\begin{align}
\Psi(u)=\lambda(1-e^{iu}).
\end{align}
Random variables can be obtained to simulate the path of assets by calling:
 \begin{verbatim}
#Poisson generator
PPgen=function(lambda){
X=0
Sum=0
flag=0
while (flag==0){
E=-log(runif(1))
Sum=Sum+E
if (Sum < lambda) { X=X+1} else { flag=1}
}
return(X)
}

#Poisson process
PP=function(lambda, N, T){
h=T/N
t=(0:T)/N
X=rep(0, N+1)
I=rep(0,N)
X[1]=0
for(i in 1:N) { 
I[i]=PPgen(h*lambda)
X[i+1]=X[i] + I[i]}
return(X)
}
\end{verbatim}
\begin{figure}[H] 
\centerline{\includegraphics[scale=0.55]{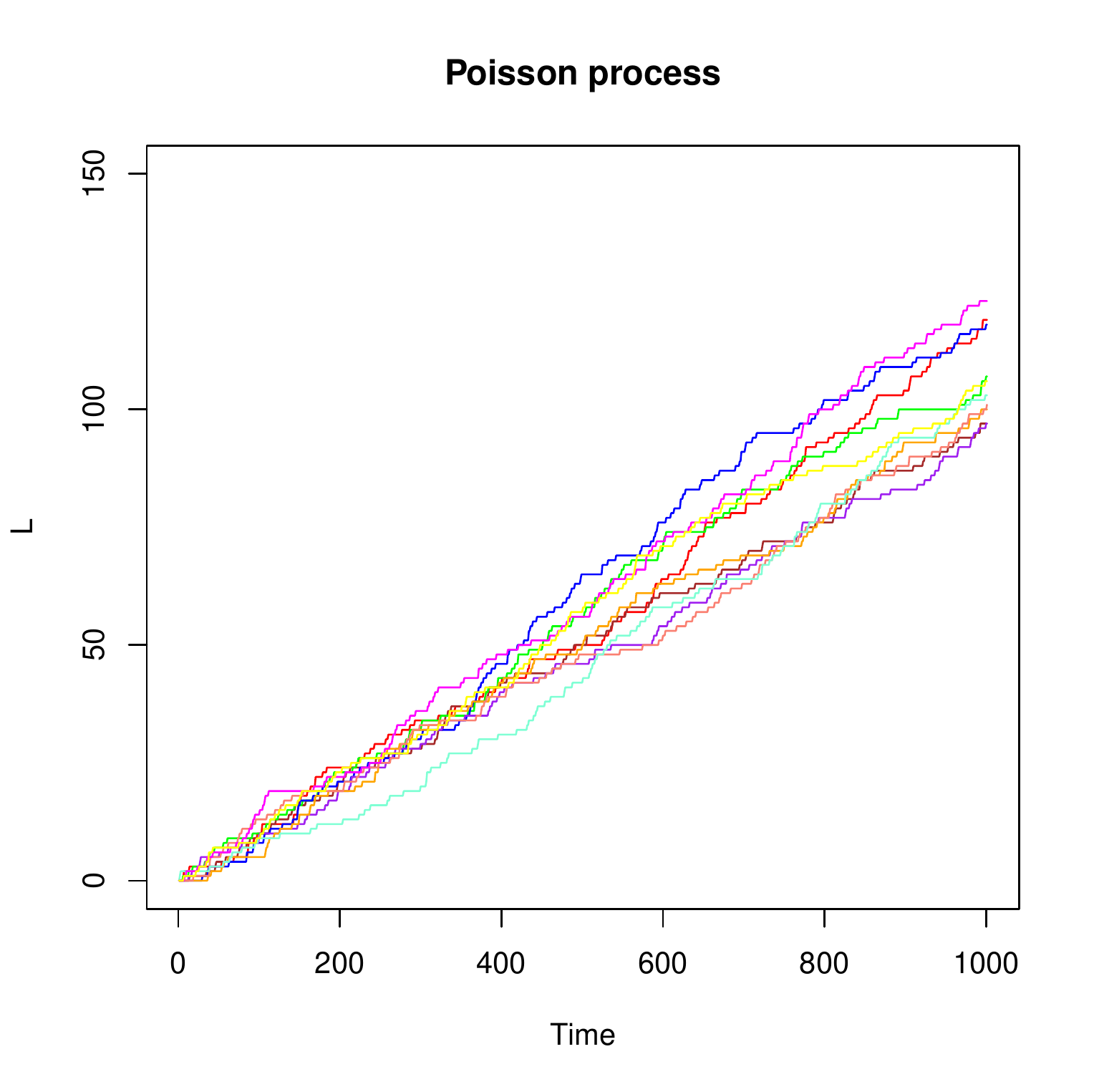}} \label{PP}
\caption{Sample paths of a Poisson process with $\lambda=100$.}
\end{figure} \noindent
\textbf{Compound Poisson Process:} Let $\lambda >0$ be the intensity of a Poisson random variable $N$ and let $\xi_i$ be random variables with law $F$ for $ i \geq 1$ (independent of N). Suppose $L_1$ follows a compound Poisson distribution. Then the triplet is given by $(-\lambda \int_{0<|x|<1} x F(dx),0, \lambda F(dx))$. The density $\eta_{L_1}$ is not analytically tractable. The canonical decomposition is given by,
\begin{align}
L_t= \displaystyle\sum_{i=1}^{N_t} \xi_i
\end{align}
which has a characteristic exponent of the form,
\begin{align}
\Psi(u)=\lambda \int\limits_\R (1-e^{iux}) F(dx).
\end{align}
Random variables can be obtained to simulate the path of assets by calling: \footnote{Note that the function PPgen is required from before.}
 \begin{verbatim}
CPP=function(lambda, T, N) {
h=T/N
t=(0:T)/N
X=rep(0, N+1)
F=rep(0, N+1)
I=rep(0,N)
X[1]=0
for(i in 1:N) { 
I[i]=PPgen(h*lambda)
if (I[i]==0){F[i]=0} else {F[i]=rnorm(1)} 
X[i+1]=X[i] + F[i]}
return(X)
}
\end{verbatim}
\begin{figure}[H] 
\centerline{\includegraphics[scale=0.55]{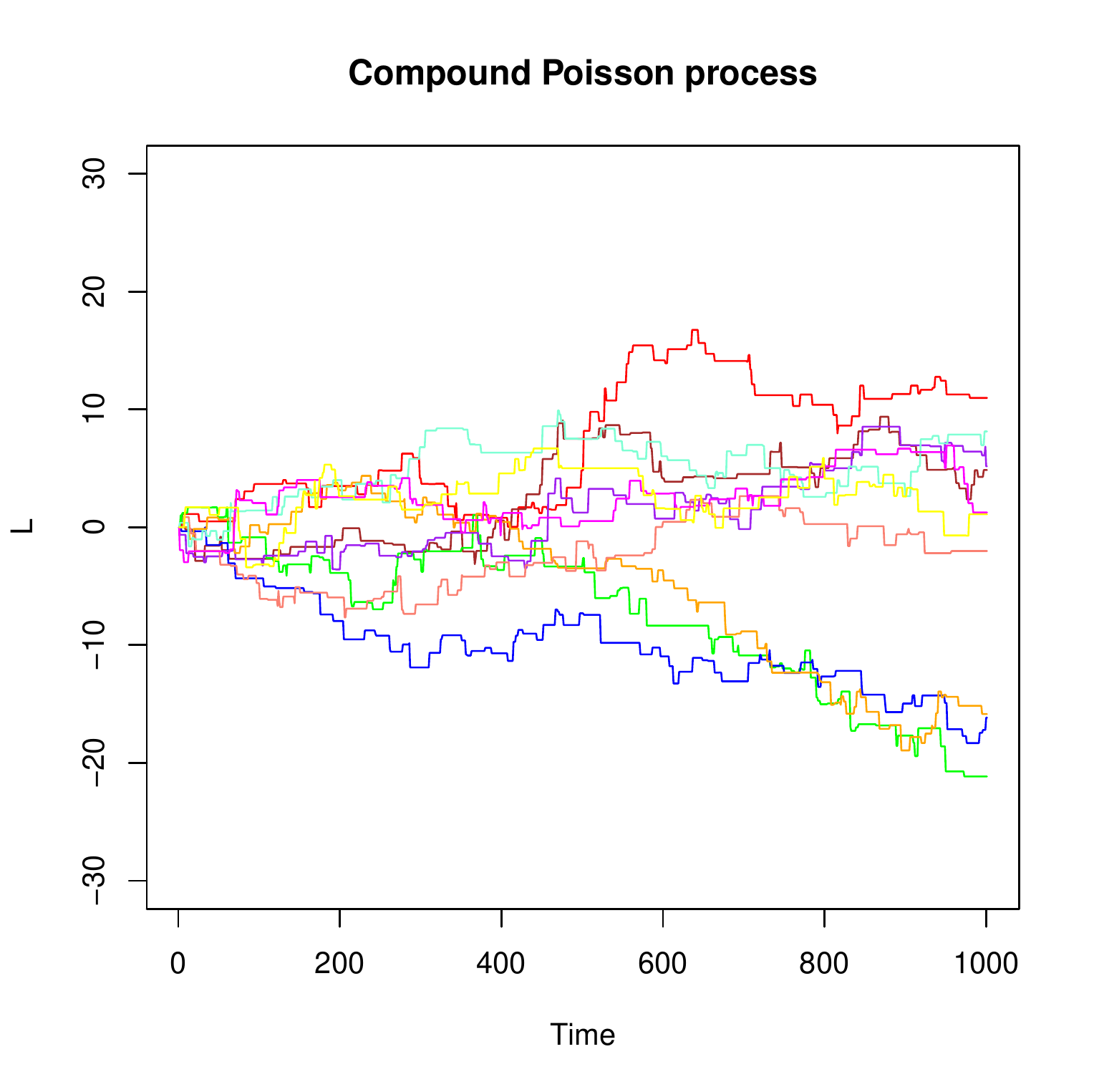}} \label{CPP}
\caption{Sample paths of a compound Poisson process with $\lambda=100$ and a standard normal jump distribution.}
\end{figure} \noindent
\textbf{Merton model:} Suppose that log-returns $L_1$ are determined by the Merton model; the sum of standard Brownian motion (with drift) and a compound Poisson process whose jumps are normally distributed (\cite{1999866}). Then $\xi_i \sim \Normal(\mu_\xi, \sigma_\xi)$ for $i \geq 1$ and the triplet is $(\mu, \sigma^2, \lambda \times \eta_\xi)$. As expected, the density $\eta_{L_1}$ is not analytically tractable. The density $\eta_{\xi}$ is given by,
\begin{align}
\eta_{\xi} (x) = \frac{1}{\sqrt{2 \pi} \sigma_\xi} e^{-(x-\mu_\xi)^2 /2\sigma_\xi^2}  .
\end{align}
The canonical decomposition is given by,
\begin{align}
L_t= \mu t + \sigma W_t + \displaystyle\sum_{i=1}^{N_t} \xi_i
\end{align}
which has a characteristic exponent of the form,
\begin{align}
\Psi(u)=\frac{1}{2} \sigma^2 u^2 - iu\mu + \lambda (1-e^{iu\mu_\xi - \sigma^2_\xi u^2 /2}).
\end{align}
which can be obtained via:
\begin{verbatim}
Merton_CF=function(u,sigma,a,b,lambda, time){
jump=lambda*time*(-a*u*1i+(exp(u*1i*log(1+a)+0.5*b*b*u*1i*(u*1i-1))-1))
phi=BS_CF(u,sigma,r,time)+exp(jump)
return(phi)
}
\end{verbatim}
Random variables can be obtained to simulate the path of assets by calling: \footnote{Note that the function PPgen is required from before.}
 \begin{verbatim}
Merton=function(mu,sigma,lambda,mu_xi,sigma_xi, N, T) {
h=T/N
t=(0:T)/N
X=rep(0, N+1)
F=rep(0, N+1)
I=rep(0,N)
X[1]=0
for(i in 1:N) { 
I[i]=PPgen(h*lambda)
if (I[i]==0){F[i]=0} else {F[i]=mu_xi*I[i]+ sqrt(sigma_xi)*sqrt(I[i])*rnorm(1)} 
X[i+1]=X[i] + mu*h+sigma*sqrt(h)*rnorm(1)+F[i]}
return(X)
}
\end{verbatim}
\begin{figure}[H] 
\centerline{\includegraphics[scale=0.55]{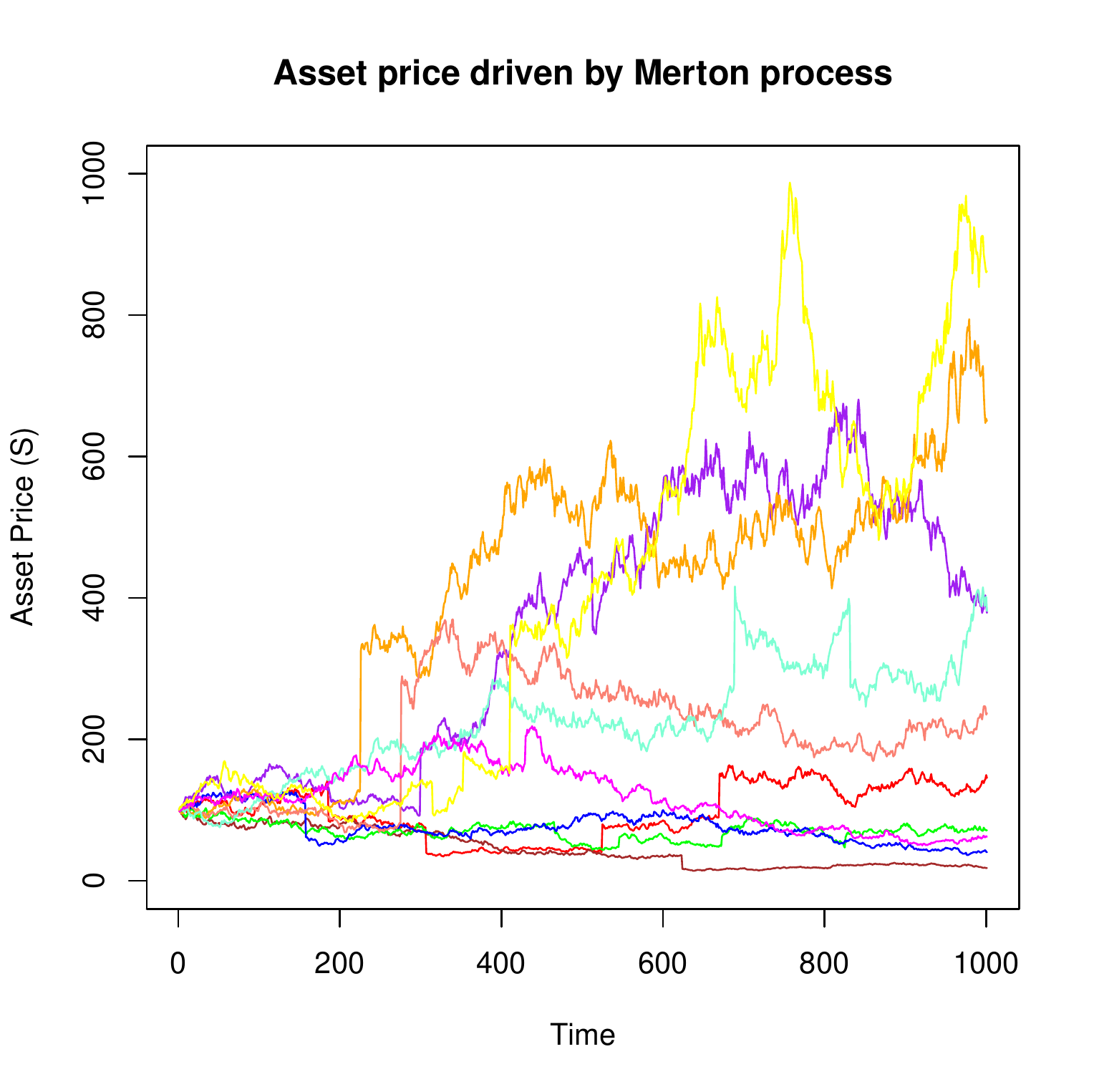}} \label{Mertonprice}
\caption{Sample paths of an asset driven by a Merton process with $\mu=0.5$, $\sigma=0.75$, and $\lambda=1.5$.}
\end{figure} \noindent
\textbf{Kou model:} Suppose that log-returns $L_1$ are determined by the Kou model; the sum of standard Brownian motion (with drift) and a compound Poisson process whose jumps are double-exponentially distributed (\cite{kou}). Then $\xi_i \sim \DExp(p,\theta_1,\theta_2)$ for $i \geq 1$ and the triplet is $(\mu, \sigma^2, \lambda \times \eta_\xi)$. Once again, the density $\eta_{L_1}$ is not analytically tractable. The density $\eta_{\xi}$ is given by,
\begin{align}
\eta_{\xi} (x) = p\theta_1 e^{-\theta_1 x} \mathds{1}_{ \{ x <0\} } + (1-p) \theta_2 e^{\theta_2 x} \mathds{1}_{ \{ x >0\} }.
\end{align}
The canonical decomposition is given by,
\begin{align}
L_t= \mu t + \sigma W_t + \displaystyle\sum_{i=1}^{N_t} \xi_i
\end{align}
which has a characteristic exponent of the form,
\begin{align}
\Psi(u)=\frac{1}{2} \sigma^2 u^2 - iu\mu + \lambda \bigg(1 + \frac{(1-p) \theta_2}{\theta_2 + iu} - \frac{p \theta_1}{ \theta_1 -iu}\bigg).
\end{align}
which can be obtained via:
\begin{verbatim}
Kou_CF=function(u,sigma, lambda, p, theta_1, theta_2, r, time){
drift=r-0.5*sigma^2 -lambda*(((p*theta_1)/(theta_2+1))+((1-p)*theta_2/(theta_2+1))-1)
phi=exp(1i*u*drift*time-0.5*sigma^2*u*u*time+time*lambda*((p*theta_1)/(theta_1+u*1i)+((1-p)*theta_2)/(theta_2+u*1i)-1))
return(phi)
}
\end{verbatim}
Random variables can be obtained to simulate the path of assets by calling: \footnote{Note that the function PPgen is required from before.}
 \begin{verbatim}
Kou=function(mu,sigma,lambda,p,theta_1,theta_2, T, N) {
h=T/N
t=(0:T)/N
X=rep(0, N+1)
F=rep(0, N+1)
I=rep(0,N)
X[1]=0
for(i in 1:N) { 
I[i]=PPgen(h*lambda)
if (I[i]==0){F[i]=0} 
else {
K=rbinom(1,I[i],p)
R1=rgamma(1,K*theta_1,theta_2)
R2=rgamma(1,(I[i]-K)*theta_1,theta_2)
F[i]=R1-R2} 
X[i+1]=X[i] + mu*h+sigma*sqrt(h)*rnorm(1)+F[i]}
return(X)
}
\end{verbatim}
\begin{figure}[H] 
\centerline{\includegraphics[scale=0.55]{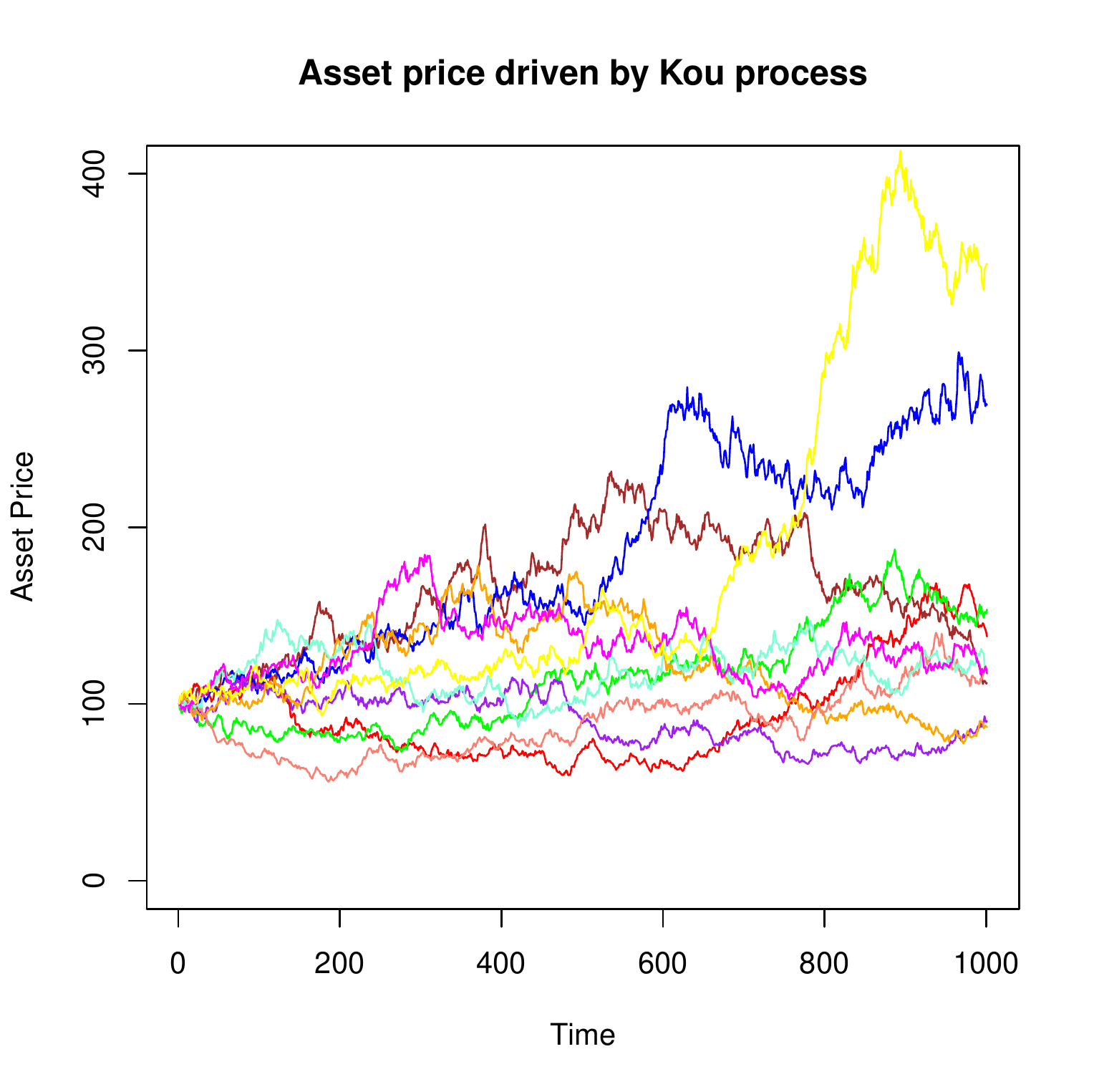}} \label{Kouprice}
\caption{Sample paths of an asset driven by a Kou process with $\mu=0.5$, $\sigma=0.75$, $\lambda=1.5$, $p=0.5$, $\theta_1=0.25$, and $\theta_2=0.25$.}
\end{figure} \noindent
{\sl Jump-diffusion } models are generated by processes with a non-zero diffusion component and finite jumps. The above examples are all of this type since $\sigma >0$ and $\nu(dx) < \infty$. If instead $\sigma = 0$ and $\nu(dy)= \infty$, we obtain the class of {\sl infinite activity pure jump} models. The majority of asset pricing models proposed in recent years are of this type. \\

\noindent \textbf{Variance Gamma Process:} Suppose that log-returns $L_1$ are determined by a variance gamma price process (\cite{madan1986chebyshev}, \cite{1990madan}). The variance gamma process can be obtained by evaluating standard Brownian motion with drift, subordinated by a gamma process in the time variable. Let $\mu, \sigma, v >0$
\begin{align} \label{cgm1}
C &=\frac{1}{v}>0 \\ \label{cgm2}
G &= \bigg(\sqrt{\frac{1}{4} \mu^2 v^2 + \frac{1}{2} \sigma^2 v} -\frac{1}{2} \mu v \bigg)^{-1} > 0\\ \label{cgm3}
M &= \bigg(\sqrt{\frac{1}{4} \mu^2 v^2 + \frac{1}{2} \sigma^2 v }+\frac{1}{2} \mu v \bigg)^{-1} > 0.
\end{align}
By defining $\gamma = \frac{C}{MG} \big[  M (e^{-G} -1) - G(e^{-M}-1) \big]$ and,
\begin{align}
\nu^{VG}(dx) =   Ce^{G|x|} |x|^{-1} \mathds{1}_{x < 0}dx +   Ce^{-Mx} x^{-1} \mathds{1}_{ x > 0}dx
\end{align}
 the triplet becomes $(\gamma, 0, \nu^{VG})$. The density $\eta_{L_1}$ is not analytically tractable (page 4, rathgeber, stadler, stockl). The canonical decomposition is given by,
\begin{align} \label{VGprocess}
L_t=  t \E[L_1]+ \int\limits_0^t \int\limits_\R x(\eta^L -\nu^{VG})(ds,dx)
\end{align}
which has a characteristic exponent of the form,
\begin{align}
\Psi(u)=\frac{1}{v} \ln \big[1-iu \mu v + \frac{1}{2} \sigma^2 v u^2 \big].
\end{align}
which can be obtained via:
\begin{verbatim}
VG_CF=function(u,sigma, theta, nu, r, time){
drift=r+log(1-theta*nu -0.5*sigma*sigma*nu)/nu
phi=exp(1i*drift*time*u) *((1-1i*nu*theta*u+0.5*nu*sigma*sigma*u*u)^(-time/nu))
return(phi)
}
\end{verbatim}
Alternatively, \eqref{VGprocess} can be expressed as the difference between two gamma processes i.e. $L_t= G_t^1-G_t^2$. The density of a gamma process is given by,
\begin{align}
\eta_{L_1}(x)= \frac{b^a}{\Gamma(a)}x^{a-1} e^{-xb}
\end{align}
where $a,b,x>0$. By setting $a=C$ and $b=M$ for $G_t^1$ and $a=C$ and $b=G$ for $G_t^2$, one can simulate the VG process in this way.  \footnote{Note that the function PPgen is required from before.}
 \begin{verbatim}
VG=function(sigma, nu, mu, T, N) {
a=1/nu
b=1/nu
h=T/N
t=(0:T)/N
X=rep(0, N+1)
I=rep(0,N)
X[1]=0
for(i in 1:N) { 
I[i]=rgamma(1,a*h,b) 
X[i+1]=X[i] + mu*I[i]+sigma*sqrt(I[i])*rnorm(1)
}
return(X)
}
\end{verbatim}
\begin{figure}[H] 
\centerline{\includegraphics[scale=0.55]{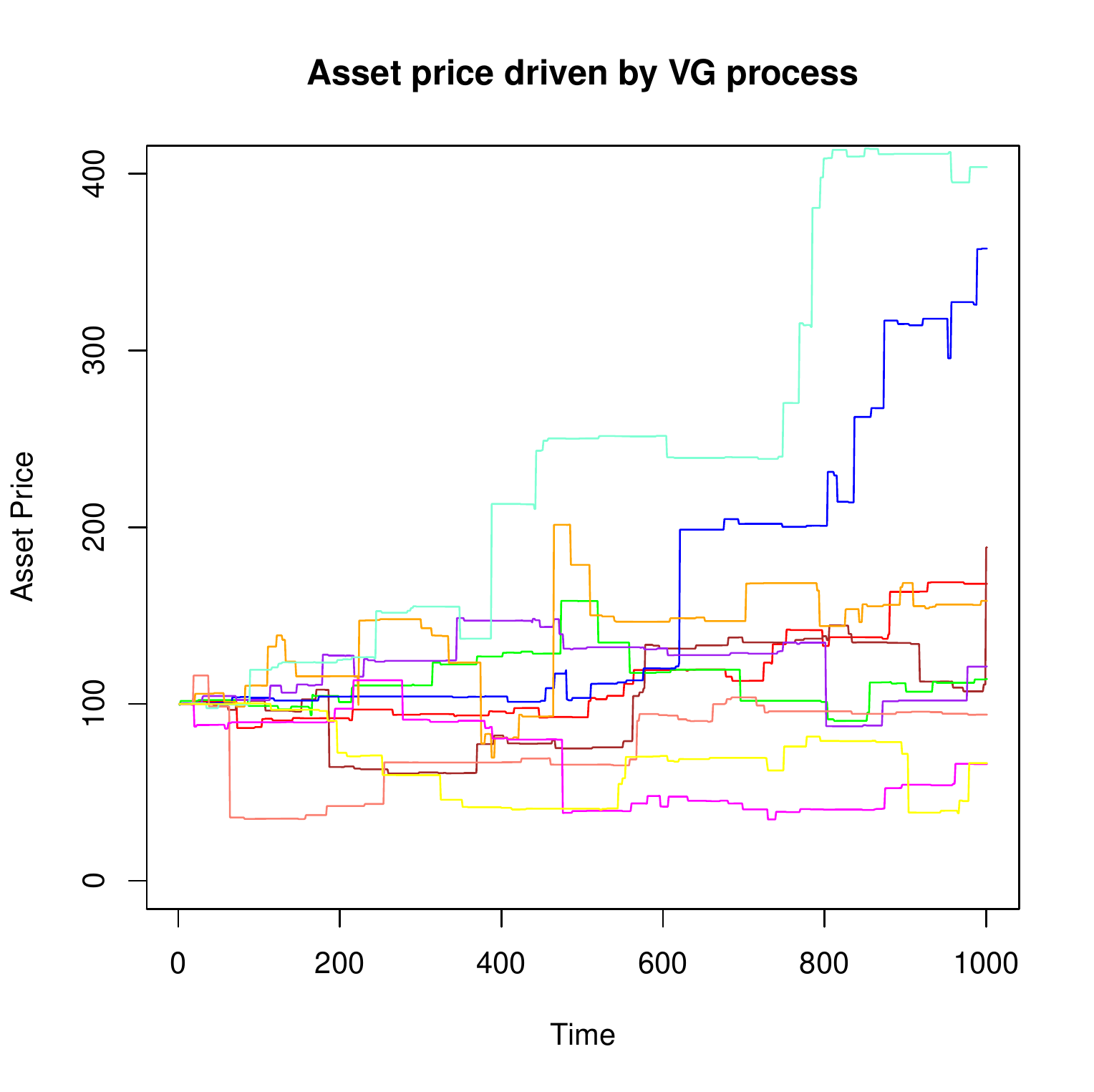}} \label{VGprice}
\caption{Sample paths of an asset driven by a VG process with $\sigma=0.75$, $v=0.5$, and $\mu=0.1$.}
\end{figure} \noindent
\textbf{CGMY Process:} Suppose that log-returns $L_1 \sim \CGMY(C,G,M,Y)$ where $C,G,M$ are adopted from \eqref{cgm1}-\eqref{cgm3} and $Y <2$ (\cite{CGMY}). \footnote{With a change of variables, this is the tempered stable process.}  This process has infinite activity iff $Y \in [0,2)$. The first moment does not have an explicit form. Let,
\begin{align}
\nu^{TS}(dx) = \frac{C e^{-Mx}}{x^{1+Y}} \mathds{1}_{x>0}dx + \frac{Ce^{-G|x|}}{|x|^{1+Y}} \mathds{1}_{x<0}dx
\end{align}
then the triplet is given by $(\E[L_1],0,\nu^{TS}(dx))$. The density $\eta_{L_1}$ is not analytically tractable. The canonical decomposition is given by,
\begin{align} \label{cgmyl}
L_t= t \E[L_1]+ \int\limits_0^t \int\limits_\R x(\eta^L -\nu^{TS})(ds,dx)
\end{align}
which has a characteristic exponent of the form,
\begin{align}
\Psi(u)=C \Gamma(-Y) [G^Y+M^Y-(M-iu)^Y-(G+iu)^Y].
\end{align}
which can be obtained via:
\begin{verbatim}
CGMY_CF=function(u,T,r,C,G,M,Y){
omega=-C*cgamma(-Y)*((M-1)^Y-M^Y+(G+1i*u)^Y-G^Y)
tmp=C*T*cgamma(-Y)*((M-1i*u)^Y-M^Y+(G+1i*u)^Y-G^Y)
phi=exp(1i*u*((r+omega)*T)+tmp)
return(phi)
}
\end{verbatim}
Similar to the VG process, the CGMY process can be simulated using Brownian subordination \cite{madanyor}. Let,
\begin{align}
A=\frac{G-M}{2} \hspace{5mm} {\text{and}}  \hspace{5mm} B=\frac{G+M}{2}.
\end{align}
Then the process defined by $L_t=AY_t+W(Y_t)$ is equivalent to \eqref{cgmyl} where $Y_t$ is the process with density,
\begin{align}
\eta_{L_1}(x)=\exp \bigg(\frac{-(B^2-A^2)y}{2} \bigg)\frac{\Gamma(Y)  h_Y (B \sqrt{y} )}{\Gamma(Y/2) 2^{Y/2-1}}
\end{align}
and for $Y>0$,
\begin{align}
h_Y(z)=\frac{1}{\Gamma(Y)} \int_0^\infty e^{-y^2/2-yz}y^{Y-1} dy.
\end{align}
Random variables can be obtained to simulate the path of assets by calling:
\begin{verbatim}
CGMY=function(C,G,M,Y, T,N) {
h=T/N
t=(0:T)/N
A=(G-M)/2
B=(G+M)/2
f=function(y){exp(-(B^2-A^2)*y/2)*cgamma(Y)*CGMY_f(y,Y,B)/(cgamma(Y/2)*2^(Y/2-1))}
X=rep(0, N+1)
I=rep(0,N)
probJ=rep(0,N)
J=(0:(N-1))*h + h
for (j in 1:N) {probJ[[j]]<-f(J[[j]])}
randf=sample(J,N,probJ,replace=TRUE)
X[1]=0
for(i in 1:N) { 
I[i]=randf[[i]]
X[i+1]=X[i] + A*I[i]+sqrt(I[i])*rnorm(1)
}
return(X)
}
\end{verbatim}
\begin{figure}[H] 
\centerline{\includegraphics[scale=0.55]{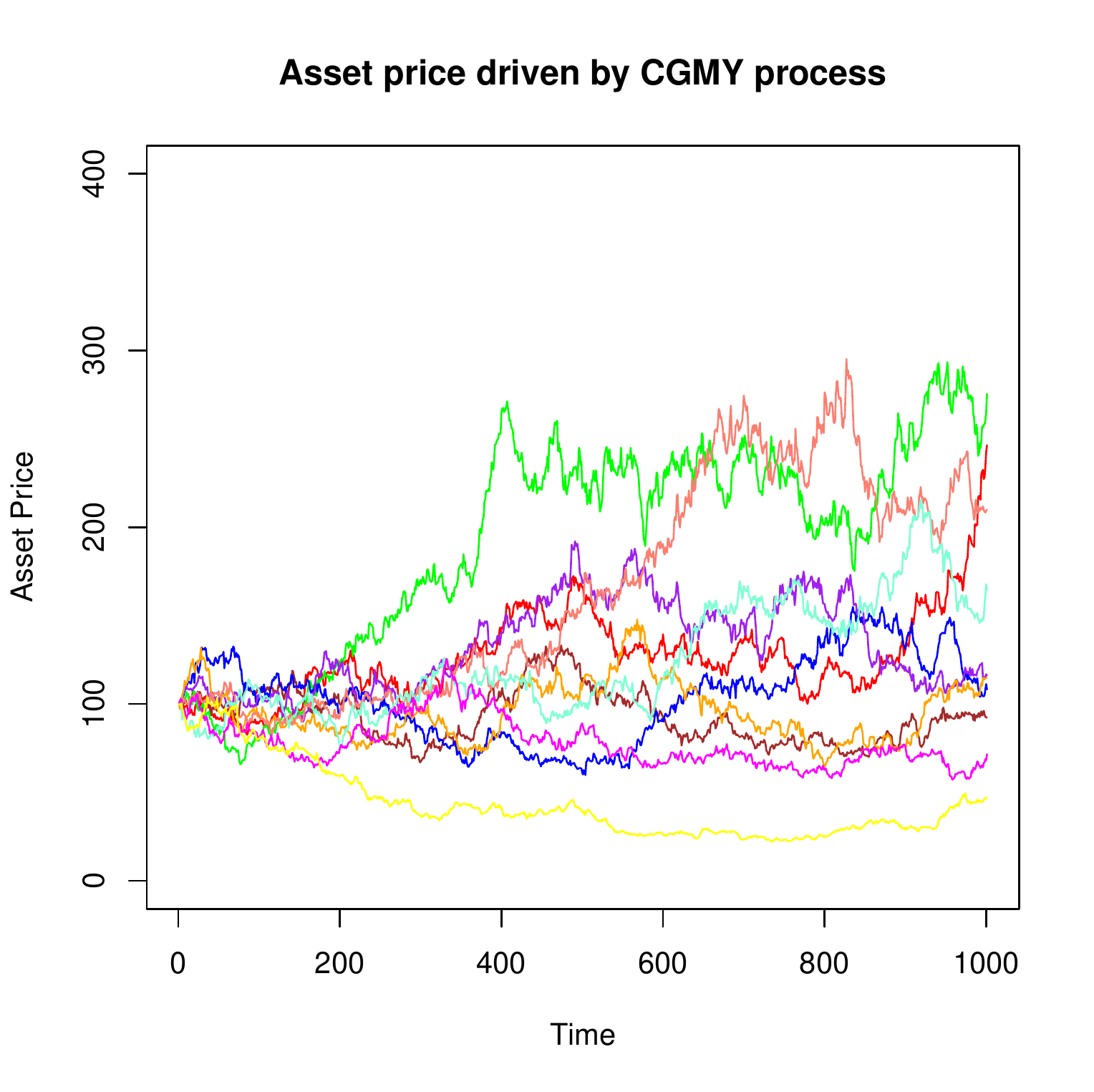}} \label{CGMYprice}
\caption{Sample paths of an asset driven by a CGMY process with $C=5$, $G=25$, $M=25$, and $Y=1$.}
\end{figure} \noindent
\textbf{Generalized Hyperbolic Process:} Suppose that log-returns $L_1 \sim \GH(\alpha, \beta, \delta, \mu, \lambda)$ (\cite{Eberlein98thegeneralized}, \cite{prause1999generalized}, \cite{barn1977}). The parameter $\alpha >0$ determines the shape, $0 \leq |\beta| < \alpha$ determines the skewness, $\mu \in \R$ gives the location, $\delta > 0$ is the scaling parameter, and $\lambda \in \R$ determines the heaviness of the tails. The first moment is given by,
\begin{align}
\E[L_1]&= \mu + \frac{\beta \delta^2 K_{\lambda+1} ( \zeta)}{\zeta K_{\lambda}(\zeta) } 
\end{align}
where $\zeta=\delta \sqrt{\alpha^2-\beta^2}$. By defining
\begin{align}
\nu^{GH}(dx)&=\frac{e^{\beta x}}{|x|} \Bigg( \int\limits_0^\infty \frac{ e^{-\sqrt{2y+\alpha^2} |x|}  }{\pi^2 y (J_{|\lambda|}^2(\delta \sqrt{2y}) + Y_{|\lambda|}^2(\delta \sqrt{2y})} dy + \lambda e^{-\alpha|x|} \mathds{1}_{ \{ \lambda \geq 0 \} }   \Bigg)
\end{align}
the triplet is given by $(\E[L_1],0,\nu^{GH}(dx))$. The density $\eta_{L_1}$ is known:
\begin{align}
\eta_{L_1}(x) = c(\alpha, \beta, \delta, \lambda)[\delta^2 + (x- \mu)^2]^{(\lambda-\frac{1}{2})/2} K_{\lambda -\frac{1}{2}}(\alpha \sqrt{\delta^2+(x-\mu)^2})e^{\beta(x-\mu)}
\end{align}
where
\begin{align}
c(\alpha, \beta, \delta, \lambda) = \frac{(\alpha^2-\beta^2)^{\frac{\lambda}{2}}}{\sqrt{2 \pi} \alpha^{\lambda-\frac{1}{2}}K_\lambda(\delta \sqrt{\alpha^2-\beta^2})}.
\end{align}
The canonical decomposition is given by,
\begin{align}
L_t= t \E[L_1]+ \int\limits_0^t \int\limits_\R x(\eta^L -\nu^{GH})(ds,dx)
\end{align}
which has a characteristic exponent of the form,
\begin{align}
\Psi(u)=-\ln \bigg[ \bigg( \frac{\alpha^2-\beta^2}{\alpha^2 - (\beta+iu)^2} \bigg)^{\frac{\lambda}{2}} \frac{K_\lambda ( \delta \sqrt{\alpha^2 -(\beta + iu)^2})}{K_\lambda (\delta \sqrt{\alpha^2-\beta^2})}    \bigg] - iu\mu.
\end{align}
which can be obtained via:
\begin{verbatim}
GH_CF=function(u,T,r,alpha, beta, delta, nu){
arg1=alpha*alpha-beta*beta
arg2=arg1-2*1i*u*beta+u*u
argm=arg1-2*beta-1
omega= - log((arg1/argm)^(0.5*nu)*BesselK(delta*sqrt(argm),nu)/BesselK(delta*sqrt(arg1), nu))
tmp=(arg1/arg2)^(0.5*nu)*BesselK(delta*sqrt(arg2), nu)/BesselK(delta*sqrt(arg1), nu)
phi=exp(1i*u*((r+omega)*T)+log(tmp)*T)
return(phi)
}
\end{verbatim}
A natural way to simulate the GH process is to directly compute,
\begin{align}
L_t=\mu+\beta Y_t+ W(Y_t).
\end{align}
where $Y_t$ is a random variable from a generalized inverse Gaussian (GIG) distribution. The density of the GIG distribution is given by,
\begin{align}
\eta_{L_1}(x)= \frac{(a/b)^{p/2}}{2 K_p (\sqrt{ab})} x^{p-1} e^{-(ax+b/x)/2}
\end{align}
where $a,b,x>0$, $p \in \R$, and $K_p$ is a modified Bessel function of the second kind. By setting $a=\alpha^2-\beta^2$, $b=\delta^2$, and $p=\lambda$ we obtain the density of $Y$. Random variables can be obtained to simulate the path of assets by calling:
\begin{verbatim}
GH=function(alpha,beta,delta,lambda,mu, T, N){
h=T/N
t=(0:T)/N
I=rep(0, N)
X=rep(0, N+1)
for (i in 1:N) {
I[i] = rgig(1, lambda,  h*sqrt(alpha*alpha-beta*beta),delta)
X[i+1]=X[i]+mu/N+beta*I[i]+ sqrt(I[i])*rnorm(1)
}
return(X)
}
\end{verbatim}
\noindent \textbf{Normal Inverse Gaussian Process:} By setting $\lambda=-\frac{1}{2}$ for the GH process, we obtain the NIG (\cite{barndorff1995normal}). The L\'evy triplet is inherited from before as $(\E[L_1],0,\nu^{NIG}(dx))$ where the first moment is given by,
\begin{align}
\E[L_1] =\mu + \frac{\beta \delta}{\sqrt{\alpha^2-\beta^2}}
\end{align}
and the measure is defined as,
\begin{align}
\nu^{NIG} (dx) = e^{\beta x} \frac{\delta \alpha}{\pi |x|} K_1 (\alpha |x|) dx.
\end{align}
Furthermore, the canonical decomposition is given similarly by, 
\begin{align}
L_t= t \E[L_1]+ \int\limits_0^t \int\limits_\R x(\eta^L -\nu^{NIG})(ds,dx)
\end{align}
which has a characteristic exponent of the form,
\begin{align}
\Psi(u)=\delta \big[ \sqrt{\alpha^2-(\beta^2+iu)^2} - \sqrt{\alpha^2 -\beta^2} \big] -iu\mu.
\end{align}
\begin{verbatim}
NIG_CF=function(u,time, r,alpha,beta,delta,mu){
omega=delta*(sqrt(alpha*alpha-(beta+1)^2)-sqrt(alpha*alpha-beta*beta))
tmp=1i*u*mu*time-delta*time*(sqrt(alpha*alpha-(beta+1i*u)^2)-sqrt(alpha*alpha-beta*beta))
phi=exp(1i*u*((r+omega)*T)+tmp)
return(phi)
}
\end{verbatim}
 Random variables can be obtained to simulate the path of assets by calling:
\begin{verbatim}
#NIG process
NIG=function(alpha,beta,delta,mu, T, N){
a=1
b=delta*sqrt(alpha*alpha-beta*beta)
h=T/N
t=(0:T)/N
I=rep(0, N)
X=rep(0, N+1)
X[1]=0
for (i in 1:N) {
I[i] = IG2(a*h, b)
X[i+1]=X[i]+mu/N+beta*delta*delta*I[i]+ delta*sqrt(I[i])*rnorm(1)
}
return(X)
}
\end{verbatim}
\begin{figure}[H] 
\centerline{\includegraphics[scale=0.55]{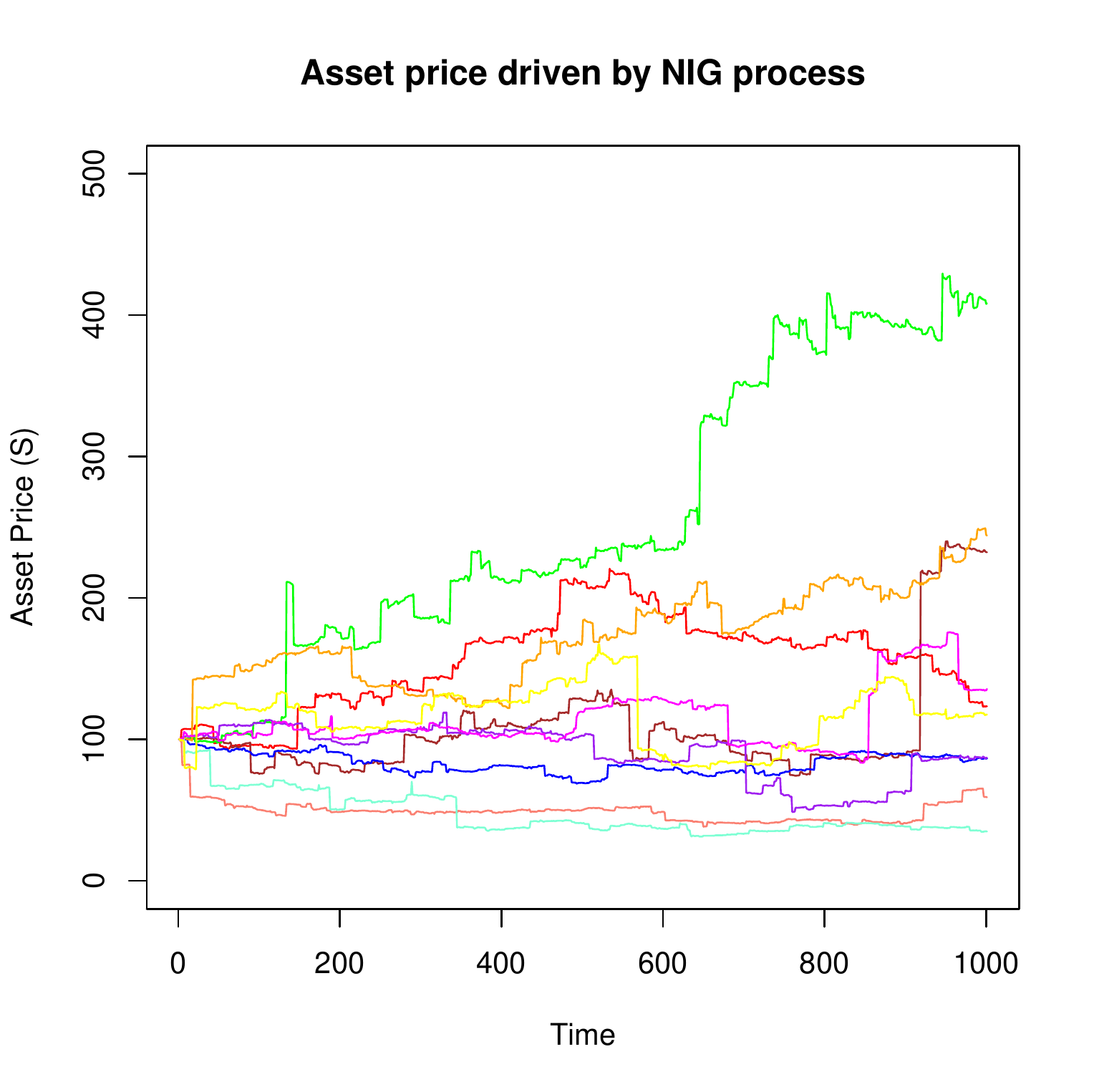}} \label{NIGprice}
\caption{Sample paths of an asset driven by a NIG process with $\alpha=2$, $\beta=1$, $\delta=1.5$, and $\mu=0$.}
\end{figure} \noindent 
\textbf{Meixner Process:} Let $\alpha,\delta >0$ and  $- \pi < \beta < \pi$ so that $L_1 \sim \MX(\alpha,\beta, \delta)$  (\cite{teugels}, \cite{schoutensmeix}). Define, 
\begin{align}
\nu^{MX}(dx) = \frac{\delta e^{\frac{\beta x}{\alpha}}}{x \sinh(\frac{ \pi x }{\alpha})}
\end{align}
then similar to the other pure jumps models, the triplet is given by $(\E[L_1],0,\nu^{MX})$. The density $\eta_{L_1}$ is,
\begin{align}
\eta_{L_1} (x) = \frac{\big(  2 \cos \big(\frac{\beta}{2} \big)\big) ^{2 \delta} }{2 \alpha \pi \Gamma(2 \delta)} \big|  \Gamma \big(  \delta + \frac{ix}{\alpha} \big) \big| e^{\frac{\beta x}{\alpha}}.
\end{align}
The canonical decomposition is given by,
\begin{align}
L_t= t \E[L_1]+ \int\limits_0^t \int\limits_\R x(\eta^L -\nu^{MX})(ds,dx)
\end{align}
which has a characteristic exponent of the form,
\begin{align}
\Psi(u)= \bigg( \frac{\cos (\frac{\beta}{2})}{\cosh( \frac{\pi x}{\alpha})}  \bigg)^{2 \delta}.
\end{align}
\begin{verbatim}
MX_CF=function(u,T,r,alpha, beta, delta){
omega=-2*delta*(log(cos(0.5*beta))-log(cos((alpha+beta)/2)))
tmp=(cos(0.5*beta)/cosh(0.5*(alpha*u-1i*beta)))^(2*T*delta)
phi=exp(1i*u*((r+omega)*T)+log(tmp))
return(phi)
}
\end{verbatim}
For simulation of the process via Brownian subordination, see \cite{madanyor}. Alternatively, for a simulation method based on acceptance-rejection sampling, see \cite{kawai}.  Random variables can be obtained to simulate the path of assets by calling: 
\begin{verbatim}
MX=function(alpha, delta, T, N) {
h=T/N
X=rep(0,N+1)
for (i in 1:N){X[[i+1]]=X[[i]]+rmeixner(h,alpha,0,delta)}
return(X)
}

#Meixner Increment Generator
rmeixner = function(t,a,m,r) {
b=0
  r=t*r
  m=t*m
  repeat {
    Q = runif(1,min=-1,max=1)/runif(1,min=-1,max=1)
    V = a/2*max(sqrt(2*r),2*r)*Q
    U = runif(1)
    if (abs(Q) < 1) {
      if (gamma(r)^2*U < abs(cgamma(r+1i*V/a))^2) { 
        break
      }  
    } 
    else {
      if (!is.na(cgamma(r+1i*V/a))) 
        if (max(1,2*r)*a^2*gamma(r+1)^2*U/(2*r) < abs(cgamma(r+1i*V/a))^2 *V^2){ 
             break
           }
    }     
  }
  return(V+m)
}
\end{verbatim}
\begin{figure}[H] 
\centerline{\includegraphics[scale=0.55]{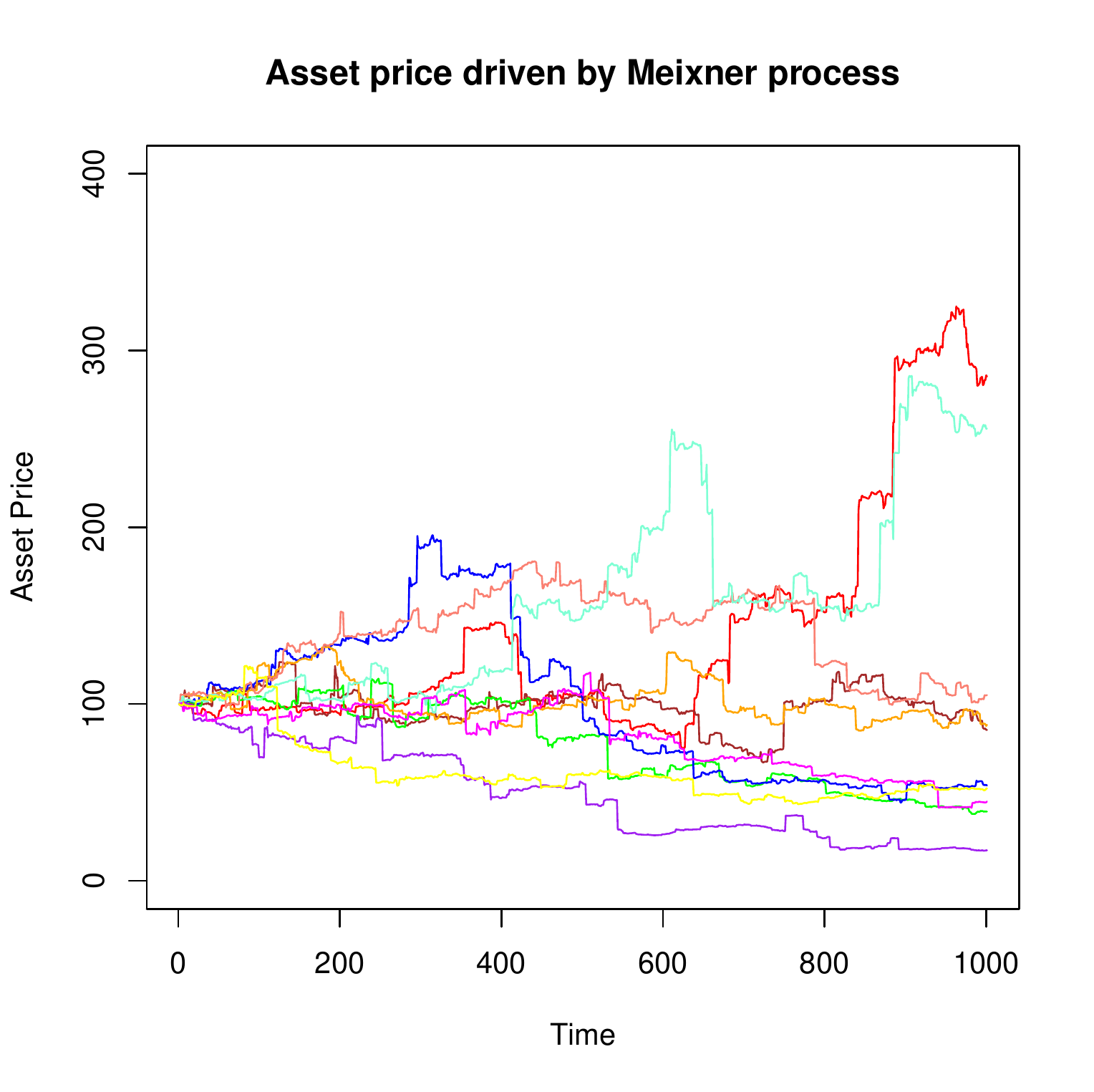}} \label{MXprice}
\caption{Sample paths of an asset driven by a MX process with $\alpha=0.5$, $\beta=0$, and $\delta=4$.}
\end{figure} \noindent
For a more detailed discussion on L\'evy processes, including their applications in finance, see \cite{kyprianou2006introductory}, \cite{protter}, \cite{papapantoleon2006}, \cite{raible2000levy}, \cite{Papapantoleaon2008}, \cite{applebaum2004levy}.\\

{\bf \noindent Should there be errors in this draft, please check with published references for the correct formulations.} \\

\bibliographystyle{plain}
\bibliography{refs}

\end{document}